\documentclass{IEEEtaes}
\usepackage{comment}
\usepackage[numbers,sort&compress]{natbib}
\usepackage[utf8]{inputenc}
\usepackage{amsmath}
\usepackage{amsfonts}
\usepackage{amssymb}
\usepackage{caption}
\usepackage{subcaption}
\usepackage{graphicx}
\usepackage{algpseudocode}
\usepackage{algorithm}
\usepackage[dvipsnames]{xcolor}
\usepackage{tikz}
\usepackage{tabularx}
\usepackage{adjustbox}
\usetikzlibrary{positioning}
\newcommand{\rece}{{{y}}}
\newcommand{\nullhypo}{{\mathcal{H}_0}}
\newcommand{\althypo}{{\mathcal{H}_1}}
\newcommand{\signal}{{{s}}}
\newcommand{\channel}{{\mathbf{H}_t}}
\newcommand{\channelt}{{{h_t}}}
\newcommand{\channelc}{{{h_c}}}
\newcommand{\clutter}{{\mathbf{H}_c}}
\newcommand{\noise}{n}
\newcommand{\Datasamples}{{n}}
\newcommand{\vecnoise}{{\mathbf{n}_s}}
\newcommand{\chanSig}{{\mathbf{y}_t}}

\newcommand{\vecrece}{{\mathbf{y}}}
\newcommand{\vecsignal}{{\mathbf{s}}}

\newcommand{\psd}{{\mathbf{R}_c}}

\newcommand{\NoiseClutter}{{\mathbf{R}}}
\newcommand{\NoiseClutterEig}{{\lambda}}
\newcommand{\noiseCoVar}{{\sigma^2\,\mathbf{I}}}
\newcommand{\clutNoise}{{\mathbf{y}_c}}
\newcommand{\EX}{{\mathbb{E}}}
\newcommand{\Complex}{{\mathbb{C}}}
\newcommand{\iid}{{\textit{i.i.d}}}
\newcommand{\BigO}{{\mathcal{O}}}
\newcommand{\tr}{{\textnormal{tr}}}

\newcommand{\I}{{\mathbf{I}}}
\newcommand{\nonLinFunc}{{\eta}}
\newcommand{\real}{{\mathbb{R}}}

\newcommand{\eigenvectoru}{{\mathbf{u}}}
\newcommand{\eigenvectorv}{{\mathbf{v}}}
\newcommand{\SampleCov}{{\hat{\textbf{R}}}}
\newcommand{\SampleCovEig}{{\hat\lambda}}
\newcommand{\Est}{{\bar{\mathbf{R}}}}
\newcommand{\EstEig}{{\bar{\lambda}}}
\newtheorem{theorem}{Theorem}

\newtheorem{definition}{Definition}

\usepackage{paralist}

\jvol{XX}
\jnum{XX}
\jmonth{XXXXX}
\paper{1234567}
\pubyear{2023}
\doiinfo{TAES.2023.Doi Number}

\begin{document}

\title{Radar Clutter Covariance Estimation: A Nonlinear Spectral Shrinkage Approach}

\author{SHASHWAT JAIN}
%\member{Fellow, IEEE}
\affil{Cornell University, Ithaca, NY, USA} 

\author{VIKRAM KRISHNAMURTHY}
\member{Fellow, IEEE}
\affil{Cornell University, Ithaca, NY, USA} 

\author{MURALIDHAR RANGASWAMY}
\member{Fellow, IEEE}
\affil{Air Force Research Laboratory}

\author{BOSUNG KANG}
\member{Member, IEEE}
\affil{University of Dayton Research
Institute, Dayton, OH, USA}
\author{SANDEEP GOGINENI}
\member{Senior Member, IEEE}
\affil{Information Systems Laboratories Inc., Dayton,
Ohio, USA}

\receiveddate{Manuscript received XXXXX 00, 0000; revised XXXXX 00, 0000; accepted XXXXX 00, 0000. }
%% \accepteddate{XXXXX XX XXXX}
%% \publisheddate{XXXXX XX XXXX}

\corresp{}

\authoraddress{}

\editor{A preliminary 4-page version of this paper is under review at ICASSP. The ICASSP submission only includes the main result and brief numerical examples. The current paper provides a detailed exposition of the problem formulation and  main results along with extensive numerical examples and discussion.}
\supplementary{}

\markboth{Jain et al.}{Radar Clutter Covariance Estimation: A Nonlinear Spectral Shrinkage Approach}
\maketitle

\begin{abstract}   In this paper, we exploit the spiked covariance structure of the clutter plus noise covariance matrix for radar signal processing. Using state-of-the-art techniques high dimensional statistics, we propose a nonlinear shrinkage-based rotation invariant spiked covariance matrix estimator. We state the convergence of the estimated spiked eigenvalues. We use a dataset generated from the high-fidelity, site-specific physics-based radar simulation software RFView to compare the proposed algorithm against the existing Rank Constrained Maximum Likelihood (RCML)-Expected Likelihood (EL) covariance estimation algorithm. We demonstrate that the computation time for the estimation by the proposed algorithm is less than the RCML-EL algorithm with identical Signal to Clutter plus Noise (SCNR) performance. We show that the proposed algorithm and the RCML-EL-based algorithm share the same optimization problem in high dimensions. We use Low-Rank Adaptive Normalized Matched Filter (LR-ANMF) detector to compute the detection probabilities for different false alarm probabilities over a range of target SNR. We present preliminary results which demonstrate the robustness of the detector against contaminating clutter discretes using the Challenge Dataset from RFView. Finally, we empirically show that the minimum variance distortionless beamformer (MVDR) error variance for the proposed algorithm is identical to the error variance resulting from the true covariance matrix.   
\end{abstract}


\section{INTRODUCTION}
Clutter plus noise covariance matrix estimation is an integral part of radar signal analysis. In a high-dimensional setting, the sample size is of the same order of magnitude as the dimension of the covariance matrix. Therefore, the sample covariance matrix is no longer a reliable estimator of the clutter plus noise covariance matrix as it becomes singular.

To mitigate such singular nature of the sample covariance matrix, we exploit the {\em spiked covariance structure} for high dimensional settings proposed in \cite{johnstone2001distribution,paul2007asymptotics,donoho2018optimal} to model the clutter plus noise covariance matrix. We propose a rotation invariant nonlinear shrinkage-based estimator to estimate the clutter plus noise covariance matrix.

 The {\em bulk} of the eigenvalues of the spiked covariance matrix are identical, corresponding to the noise component of clutter plus noise covariance matrix. A finite number of {\em spiked} eigenvalues significantly exceed the {\em bulk} eigenvalues in magnitude, accounting for the clutter component of the clutter plus noise covariance matrix.

We model the clutter in the Challenge Dataset simulated by $\textnormal{RFView}^{\circledR}$ \cite{gogineni2022high,RFView} as a {\em spiked covariance} structure. RFView is a high-fidelity, site-specific, physics based M$\&$S tool, which enables a real time instantiation of the RF environment. This has been extensively vetted using measured data from VHF to X band with one case documented in \cite{gogineni2022high}. %{\color{blue}{Additional Info about RFView, to be added here. Reviewer might ask why RFView, there are other open source software in the market.}}

\begin{figure}[h!]
    \centering
  \includegraphics[width=\linewidth]{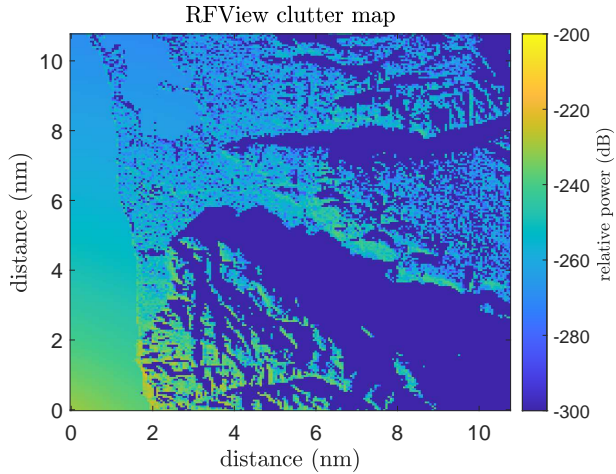}
    \caption{Clutter returns from the littoral scene with mountains and water showing shadow regions (dark blue) as well as stronger signal components (yellow).}
    \label{fig:CofarFigure}
\end{figure}

\begin{figure}[h!]
        \includegraphics[width=\linewidth,height=6.5cm]{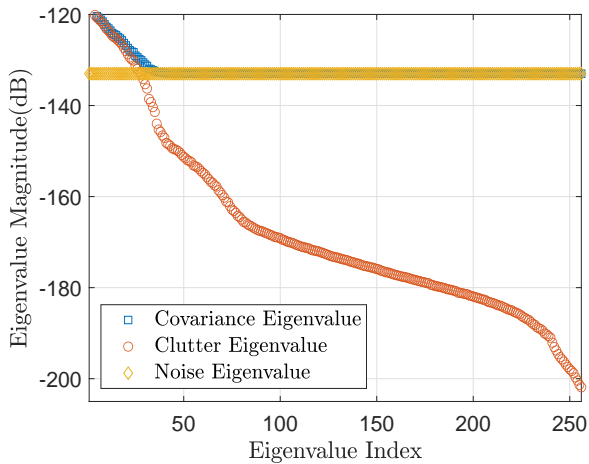}
        \caption{The clutter plus noise covariance matrix formed by the return signal given in Fig.\,\ref{fig:CofarFigure} gives evidence that most of the components of clutter (red) are below the noise floor. The first 25 components are above the noise floor. Therefore, the covariance matrix exhibits a spiked covariance structure.}
        \label{fig:ClutterNoiseCovar}
    \end{figure}

As an illustrative example, consider an airborne radar looking down on a heterogeneous terrain, consisting of mountains, water bodies, and foliage simulated by RFView as shown in Fig.\ref{fig:CofarFigure}. Fig.\ref{fig:CofarFigure} displays the relative power of the returned signal from such heterogeneous terrain in Southern California near San Diego. We observe that the regions of high-power returns have less area compared to the regions of low-power return, with noise power higher than the low-power returns.  This is evident from the eigenvalue plot of the clutter plus noise covariance matrix in Fig.\,\ref{fig:ClutterNoiseCovar}, computed from the return signal. A large number of eigenvalues of the clutter covariance matrix fall below the noise power. %This shows that for such a scenario, the clutter matrix is {\em spiked}.
    
In this paper, we use a nonlinear shrinkage-based rotation invariant estimator developed in \cite{donoho2018optimal} to estimate the clutter plus noise covariance matrix in a high dimensional setting. In the spiked covariance model, we only need to estimate the noise power and the spiked components.

We show that for the estimation of covariance matrix of dimension $p$, the proposed algorithm performs $\BigO(p)$ real-valued multiplications for the joint noise power-clutter rank estimation as compared to the $\BigO(p^2)$ in the RCML-EL\footnote{RCML-ELrepresents the RCML covariance estimator \cite{kang2014rank} with the clutter rank and noise power obtained by the expected likelihood (EL) approach \cite{kang2016expected}.} algorithm with identical SCNR and error variance. Additionally, we state the convergence results for the estimated eigenvalues and bounds for normalized SCNR for the proposed estimator. We test the target detection performance of the estimator using Low-Rank Adaptive Normalized Matched Filter (LR-ANMF) detector. We empirically show the robustness of the detector against contaminating clutter discretes. We apply the proposed algorithm on the Challenge Dataset simulated by RFView software.

\subsection{Related Works}
The problem of covariance matrix estimation, \cite{ReedMallet}, with data deficient scenario has received considerable attention in the radar signal processing literature. In the data deficient scenarios, the sample covariance matrix is no longer a reliable estimator as it becomes ill-conditioned. To address such ill-conditioning, methods like diagonal loading\,\cite{abramovich1981analysis,johnson2007matrix,abramovich1981controlled,carlson1988covariance,wicks2006space,gini2008knowledge,Ward1994} and 
factored space-time approaches\,\cite{dipietro1992extended} have been proposed. Data dependent techniques include Principal Components Inverse\,\cite{kirsteins1994adaptive}, Multistage Wiener Filter\,\cite{goldstein1998multistage}, Parametric Adaptive Matched Filter\,\cite{roman2000parametric},
and EigenCanceler\,\cite{haimovich1996eigencanceler}. Data independent approaches include JDL-GLR\,\cite{wang1994adaptive}.

In a high-dimensional setting, the properties of covariance matrices are explained by the Random Matrix Theory as stated in \cite{wainwright2019high,vershynin2018high,bai2010spectral,wang2017asymptotics}. In such high dimensional settings, shrinkage estimators have been developed to estimate covariance matrices in signal processing and finance. Shrinkage estimators have been used in wireless communications\,\cite{tulino2004random} to estimate the channel matrices, in array signal processing to estimate direction of arrival\,\cite{couillet2013joint} and in finance for Markowitz Portfolio optimization\,\cite{ledoit2018optimal,ledoit2022quadratic,ledoit2017nonlinear,ledoit2022power}. Shrinkage methods include Ledoit-Wolf shrinkage estimator\,\cite{ledoit2004well}, regularized PCA\,\cite{bun2016rotational}, Ridge and Lasso shrinkage estimators\,\cite{yuan2020improved} and regularized M-estimators\,\cite{couillet2014robust}.    

In radar signal processing, the covariance matrices often contains a low-rank structure corresponding to the clutter. The covariance matrices are called clutter plus noise covariance matrices. Covariance estimation algorithms developed in   \cite{sen2015low,spClut1,kang2016expected,kang2013constrained,abrahamsson2007enhanced}, propose estimation schemes assuming a rank sparse clutter covariance matrix in a high dimensional setting. These papers use Brennan's Rule which gives an estimate of the rank depending on the dominant components of the clutter and the jammers. However, as demonstrated in \cite{guerci2010cognitive,kang2022adaptive,guerci2016new}, Brennan's rule fails when a plethora of real-world effects such as internal clutter motion, mutual coupling between antenna array elements arise on account of the system and environmental factors. In our paper, the data generated from the high fidelity, site-specific, physics-based, radar scenario simulation software RFview is used where the Brennan rule does not prevail as documented in \cite{gogineni2022high}. 

To address these issues we exploit the {\textit{spiked covariance}} structure, as proposed in \cite{johnstone2001distribution,paul2007asymptotics,nadler2008finite,benaych2011eigenvalues}, of the clutter plus noise covariance matrix.  We use the  nonlinear shrinkage estimation techniques of \cite{donoho2018optimal} to estimate the covariance matrix. Spiked covariance models have been used to estimate direction of arrival (DOA) in array signal processing, as demonstrated in \cite{yang2018high}, and for target detection in \cite{robinson2021space}. We are using the spiked covariance model to estimate the clutter plus noise covariance matrix.

\subsection{Main Results and Organization}

The main results and organization of this paper are as follows:
\begin{compactenum}
\item In Sec.~\ref{Sec:SignalModel}-\ref{Sec:Beamforming}, we formulate the rotation invariant estimators.  The asymptotic model for large random matrices and the spiked covariance model for the clutter plus noise covariance matrix is defined in Sec.~\ref{Sec:SignalModel}-\ref{Sec:SpikedCovar}.
\item In Sec.~\ref{Sec:Donohue}, we present the algorithm proposed in \cite{donoho2018optimal} for spiked covariance matrix estimation. In Sec.~\ref{Sec:Donohue}-\ref{Sec:Thm},  Theorem\,\ref{Thm:NonLinBias} shows a strong law of large numbers (namely,  the estimated spiked eigenvalues converge almost surely to a constant) and satisfy a Central Limit Theorem. This is due to the fact that, even though we are in a high-dimensional setting,  the number of spikes are constant. Empirical verification of the convergence properties for RFview Challenge Dataset is provided.
 We derive the bounds for normalized SCNR, $\rho$, in Sec.~\ref{Sec:Donohue}-\ref{Sec:rhobound}.
In Sec.~\ref{Sec:Donohue}-\ref{Sec:Similar}, we establish that the proposed algorithm and the RCML-EL estimation algorithm in high dimensions have similar performance due to the fact they share a common optimization problem.
 In Sec.~\ref{Sec:Donohue}-\ref{Sec:Detection}, we employ the LR-ANMF detector for target detection when using a rotation invariant estimator.

\item In Sec.~\ref{Sec:Simulations}-\ref{Sec:Challenge}, we demonstrate that the proposed algorithm has identical SCNR compared to the RCML-EL algorithm for the Challenge Dataset simulated using RFView. We further show that the computation time of estimation by the proposed algorithm is less than that of the RCML-EL algorithm.
In Sec.~\ref{Sec:Simulations}-\ref{Sec:DetectionNumerical}, we compute the target detection probabilities for various false alarm probabilities and SCNR and empirically show its robustness with respect to contaminating clutter discretes. 
In Sec.~\ref{Sec:Simulations}-\ref{Sec:ErrorVariance}, we compute the error variance of the MVDR beamformer for the proposed algorithm.

\end{compactenum}

\section{Rotation Invariant Estimator and Spiked Covariance Model}
\label{Sec:SignalModel}
 This section is organized as follows. Sec.~\ref{Sec:SignalModel}-\ref{Sec:Beamforming}  presents a general rotation invariant estimator. Sec.~\ref{Sec:SignalModel}-\ref{Sec:SpikedCovar} describes the high-dimensional spiked covariance model.

We use a narrowband baseband equivalent model used in \cite{gogineni2022high}. The radar transmits a complex-valued waveform $\signal(k)\,\in\Complex$ and receives a complex-valued return $\rece(k)\,\in\Complex$ in discrete time:
\begin{equation}
\label{eq:SignalModel}
    \rece(k)=\channelt(k)\,\circledast\,\signal(k)+\channelc(k)\,\circledast\,\signal(k)+\noise(k)
\end{equation}
Here $\circledast$ is the convolution operator and $k$ denotes discrete time. $\channelt\,\in\Complex$  is the complex valued target impulse response and $\channelc\,\in\Complex$ is the complex valued clutter impulse response. The noisy measurement $\noise\,\in\Complex$ is the additive white Gaussian noise with variance $\sigma^2$ with zero mean. The noise samples are independent, identically, distributed ($\iid$).
 
In matrix-vector notation~\eqref{eq:SignalModel} reads
\begin{equation}
    \label{eq:VecSignalModel}
    \vecrece=\channel\,\vecsignal+\clutter\,\vecsignal+\vecnoise
\end{equation}
where $\channel,\,\clutter\in\Complex^{p\times q}$ are  Toeplitz matrices constructed by the impulse responses $\channelt(k)$ and $\channelc(k)$, respectively. $\vecrece\in\Complex^{p}$ is return signal of length $p$ and $\vecsignal\in\Complex^{q}$ is the waveform of pulse length $q$. The noise $\vecnoise\in\Complex^{p}$ is a complex valued Gaussian distributed vector where $\vecnoise~\sim~\mathcal{N}(\mathbf{0},\noiseCoVar)$ and has $\iid$ samples.

We define the clutter plus noise return as $\clutNoise$
\begin{equation}
    \label{eq:ChanSig}
    \clutNoise:=\clutter\,\vecsignal+\vecnoise
\end{equation}

%\vspace{-0.75cm}
\subsection{Rotation Invariant Estimator}
\label{Sec:Beamforming}
In this subsection, we describe the rotation invariant estimation for the clutter plus noise covariance estimator. Rotation invariant estimators have the same eigenvectors as that of the sample covariance matrix and the eigenvalues of the estimators are a function of the eigenvalues of the sample covariance matrix. 

The clutter plus noise covariance matrix is given by
\begin{equation}
\label{eq:Covariance}
    \NoiseClutter=\psd+\noiseCoVar
\end{equation}

where the clutter covariance matrix is
\begin{equation}
\label{eq:Clutter}
    \psd:=\EX[\clutter\,\vecsignal\vecsignal^{H}\,\clutter^{H}]
\end{equation} and $\noiseCoVar$ is the noise component. The eigendecomposition of clutter plus noise covariance is:
\begin{equation}
\label{eq:EigenDecomTrue}\NoiseClutter=\sum_{i=1}^{p}\NoiseClutterEig_{i}\eigenvectoru_i\eigenvectoru_i^{H}
\end{equation}
with eigenvalues $\NoiseClutterEig_i$ and eigenvectors $\eigenvectoru_i$.
The sample covariance matrix $\SampleCov_\Datasamples$ is:
\begin{equation}
    \label{eq:EmpiricalR}
    \SampleCov_{n}=\frac{1}{\Datasamples}\,\sum_{k=1}^{\Datasamples}\clutNoise_{,k}\,\clutNoise_{,k}^H
\end{equation}
where $\clutNoise$ is the clutter plus noise return defined in \eqref{eq:ChanSig} which will be used as training data samples\footnote{The training data is collected by RFView when no target is present. We assume that the clutter plus noise covariance matrix is stationary and is independent of  the target presence. This is due to the fact that eigenvalue component due to the target is independent of the eigenvalues of the clutter plus noise covariance matrix in the spiked covariance model which is defined in Sec.~\ref{Sec:SignalModel}-\ref{Sec:SpikedCovar}.}, $k$ is the discrete time and $\Datasamples$ are the number of training data samples. 
The spectral decomposition of $\SampleCov_{\Datasamples}$ for a given training data size $\Datasamples$ is:
\begin{equation}
\label{eq:EigenDecompSample}
    \SampleCov_{\Datasamples}=\sum_{i=1}^{p}\,\SampleCovEig_{i,n}\eigenvectorv_{i,\Datasamples}\,\eigenvectorv_{i,\Datasamples}^{H}
\end{equation}
where $\SampleCovEig_{i,\Datasamples}$ are the eigenvalues and $\eigenvectorv_{i,\Datasamples}$ are the eigenvectors of $\SampleCov_\Datasamples$. The spiked covariance matrix estimate for a given number of data samples $\Datasamples$ is:
\begin{equation}
\label{eq:EigenDecompEst}
    \Est_{\Datasamples}=\sum_{i=1}^{p}\,\EstEig_{i,n}\,\eigenvectorv_{i,\Datasamples}\,\eigenvectorv_{i,\Datasamples}^{H}
\end{equation}
where $\EstEig_{i,\Datasamples}$ are the eigenvalues of $\Est_\Datasamples$, with eigenvectors $\eigenvectorv_{i,\Datasamples}$ identical to those of the $\SampleCov_{\Datasamples}$ in \eqref{eq:EigenDecompSample}. The spiked covariance estimator is a rotation invariant estimator. 

Additionally, we use normalized SCNR to compare covariance estimation methods. We denote $\rho$ to define the normalized SCNR as: 
\begin{equation}
\label{eq:rho}
    \rho=\frac{(\chanSig^{H}\,\Est^{-1}\,\chanSig)^2}{(\chanSig^{H}\,{\NoiseClutter}^{-1}\,\chanSig)(\chanSig^{H}\,\Est^{-1}\,\NoiseClutter\,\Est^{-1}\,\chanSig)}.
\end{equation}
where $\chanSig=A_{\theta}\otimes\,A_{f}$ is the Kronecker product of angle steering vector $[A_{\theta}]_{i}=\exp[-j\pi i \sin(\theta)],\,1\leq i\leq N$ and the Doppler steering vector $[A_{f}]_{i}=\exp[-j2\pi i f_{d}], 1\leq i\leq K$. $N$ and $K$ are defined in Sec.~\ref{Sec:Simulations}. The dimension of the covariance matrix is $p=N\times K$.

In the next section, we will show that $\EstEig$ is a nonlinear function of $\SampleCovEig$, where the nonlinearity depends on the loss function.
%\vspace{-0.3cm}
\subsection{Clutter Plus Noise Covariance Matrix Modelling using Large Random Matrices}
\label{Sec:SpikedCovar}
In this subsection, we define the spiked covariance model and the asymptotic regime for the high dimensional setting. We use this framework to model the clutter plus noise covariance matrix.
%\vspace{-0.3cm}
\begin{definition}
\label{Def:SpikedCovariance}
A spiked covariance matrix $\NoiseClutter$ is a $p\times p$ positive definite Hermitian matrix with eigenvalues $(\NoiseClutterEig_{1},\,\NoiseClutterEig_{2}\,\cdots,\,\NoiseClutterEig_{p})$ such that for a finite $r\ll p$, $\NoiseClutterEig_{1}\geq\NoiseClutterEig_{2}\geq\NoiseClutterEig_{r}>\sigma^2$ and $\NoiseClutterEig_{r+1}=\cdots=\NoiseClutterEig_{p}=\sigma^2>0$. 
\end{definition}
We make two assumptions
\begin{compactenum}
\item \label{Sec:Clutter} The clutter plus noise covariance matrix has a spiked covariance structure given in Definition\,\ref{Def:SpikedCovariance}. The clutter plus noise covariance matrix is given in \eqref{eq:Covariance} where clutter covariance matrix $\psd$ has rank $r$ with eigenvalues $\NoiseClutterEig_i-\sigma^2$, $1\leq i\leq r$. The noise covariance matrix $\noiseCoVar$ is diagonal.
\item \label{Sec:Asym}There exists a $\gamma\in(0,1)$ such that for given  training data size $\Datasamples$ with the dimension of the covariance matrix as $p$ such that:
\begin{equation}
\label{eq:Kolmogorov}
    \frac{p}{\Datasamples}\rightarrow \gamma,\quad p,\,\Datasamples\rightarrow\infty,\quad p<\Datasamples
\end{equation} 
\end{compactenum}
Data displayed in Fig.\,\ref{fig:CofarFigure} and Fig.\,\ref{fig:ClutterNoiseCovar} satisfies these conditions. In Fig.\,\ref{fig:ClutterNoiseCovar}, we see that the clutter covariance matrix can be approximated by a rank $r$ positive semi-definite matrix as the remaining $p-r$ components are below the noise floor. 
%We assume that a clutter matrix is {\em spiked} only if its rank is less than ten percent of the dimension of the covariance matrix as illustrated in Fig.\,\ref{fig:ClutterNoiseCovar}. 

We assume that clutter plus noise covariance matrix is
spiked if the rank of the clutter matrix is less than a fraction $\chi$ of the clutter plus noise covariance matrix. For convenience, we choose $\chi= 0.1$, since it empirically fits with the data simulated by RFView. A more general approach involves model order (dimension) estimation. In the classical statistical setting, this is well studied in terms of penalized likelihood methods such as Akaike Information Criterion (AIC)\,\cite{akaike1974new}, Minimum Description Length (MDL)\,\cite{grunwald2007minimum}, information theoretic criteria\,\cite{wax1985detection}, statistical techniques\,\cite{bai1989rates}, data dependent techniques\,\cite{shah1994determination}, and min-max approaches such as the Embedded Exponential Families\,\cite{kay2005exponentially}. However, in the high dimensional setting considered
in this paper, estimating the model order (number of spikes) is a difficult problem not addressed in this paper. In \cite{kang2016expected}, the RCML-EL algorithm uses Brennan's rule,\,\cite{ReedMallet}, as an initial estimate for the rank of the clutter covariance matrix to determine the model order. RCML-EL algorithm correctly estimates the rank as compared to the AIC and MDL techniques. In Sec.~\ref{Sec:Donohue}-\ref{Sec:Similar}, since the proposed algorithm and RCML-EL algorithm share similar optimization problem, the proposed algorithm correctly estimates the model order.

The spiked covariance property helps us to deal with the clutter plus noise covariance matrices in high dimensions, which is frequently encountered in radar signal processing. With this knowledge, we define the nonlinear shrinkage-based rotation invariant estimator.

\section{Nonlinear Shrinkage Estimation}
\label{Sec:Donohue}
In this section, we propose the rotation invariant estimator using nonlinear shrinkage of the eigenvalues of the sample covariance matrix. We state the convergence of the estimated eigenvalues in Theorem\,\ref{Thm:NonLinBias} in Sec.~\ref{Sec:Donohue}-\ref{Sec:Thm}. It is to be noted that we use the terms spiked eigenvalues and the leading $r$ eigenvalues of the covariance matrix interchangeably for a fixed clutter covariance matrix with rank $r$. 
We outline the computation cost of the proposed algorithm. We propose bounds for the normalized SCNR($\rho$) in Sec.~\ref{Sec:Donohue}-\ref{Sec:rhobound}. We show the similarity of SCNR performance between the proposed algorithm and the RCML-EL algorithm in high dimensions in Sec.~\ref{Sec:Donohue}-\ref{Sec:Similar}. We conclude this section by stating the  Adaptive Normalized Matched Filter for target detection for the proposed algorithm in Sec.~\ref{Sec:Donohue}-\ref{Sec:Detection}.

\label{Sec:Shrinkage}
\label{Sec:ShrinkageEst}

The spiked covariance matrix $\NoiseClutter$ is stated in Definition~\ref{Def:SpikedCovariance}. 
Estimation of the spiked covariance matrix $\Est_\Datasamples$ as defined in \eqref{eq:EigenDecompEst}, consists of two sub-problems: estimation of the spiked eigenvalues and the estimation of the noise power $\sigma^2$.
\begin{enumerate}
    \item 
The estimate of the noise power is given as stated in \cite{donoho2018optimal} is
\begin{equation}
\label{eq:NoisePower}
    \hat{\sigma}^2=\frac{\SampleCovEig_{med}}{\mu_{med}}
\end{equation}
where $\SampleCovEig_{med}$ is the median of the eigenvalues of the sample covariance matrix $\SampleCov_\Datasamples$ and $\mu_{med}$ is the median of the Marchenko-Pastur distribution with parameter $\gamma$ stated in \eqref{eq:Kolmogorov}. The proof of the consistency of the noise power estimator is given in \cite[Sec.\,9]{donoho2018optimal}.\\
\item The shrinkage function $\nonLinFunc^{*}(\cdot)$ as stated in \cite{donoho2018optimal} is
\begin{equation}
\label{eq:GeneralShrinker1}
            \nonLinFunc^{*}(\Tilde\lambda_i)=\begin{cases}
            \nonLinFunc(f(\Tilde\lambda_i))& \Tilde\lambda_i>(1+\sqrt{\gamma})^{2}\\
            1 & \Tilde\lambda_i\leq(1+\sqrt{\gamma})^{2}
            \end{cases}
        \end{equation}
where $\Tilde\lambda_i=\SampleCovEig_i/\hat\sigma^2$, $\SampleCovEig_i$ are the eigenvalues of the sample covariance matrix $\SampleCov$ and $\hat\sigma^2$ is defined in \eqref{eq:NoisePower}. The function $f(\cdot)$ given by 
\begin{equation}
\label{eq:fx}
    \footnote{It is to be noted that \eqref{eq:fx} is the inverse mapping for \begin{equation}
\label{eq:gx}
    g(x)=\begin{cases}x+\frac{\gamma\,x}{x-1}& x> 1+\sqrt{\gamma}\\
    (1+\sqrt{\gamma})^{2}&1\leq x\leq1+\sqrt{\gamma}
    \end{cases}
\end{equation} when $x>(1+\sqrt{\gamma})^{2}$, it is the relationship between the $\SampleCovEig$ and $\NoiseClutterEig$ and has been explained in \cite{donoho2018optimal}.}f(x)=\frac{x+1-\gamma+\sqrt{(x+1-\gamma)^2-4x}}{2}
\end{equation}
and $\nonLinFunc(\cdot)$ for Stein loss as stated in \cite{donoho2018optimal}, $\text{L}^{St}=\tr(\NoiseClutter^{-1}\,\Est-\I)-\log\det(\NoiseClutter^{-1}\,\Est)$, is given by
\begin{equation}
\label{eq:Shrinker1}
  \nonLinFunc^{\text{St}}(x)=\frac{x}{c(x)^2+s(x)^2\,x}  
\end{equation}
where $c(\cdot)$ is given by 
\begin{equation}
\label{eq:Cosine}
    c(x)=\begin{cases}
    \sqrt{\frac{1-\gamma/(x-1)^2}{1+\gamma/(x-1)}}&x> 1+\sqrt{\gamma}\\
    0&x\leq1+\sqrt{\gamma}
    \end{cases}
\end{equation}
$s(\cdot)^2={1-c(\cdot)^2}$ and $\gamma$ is given in \eqref{eq:Kolmogorov}.\\
The eigenvalues $\EstEig_{i,\Datasamples}$ of the estimator $\Est_{\Datasamples}$ are given by:
\begin{equation}
\label{eq:EstEig}
    \EstEig_{i,\Datasamples}=\hat\sigma^{2}\,\eta^{*}(\Tilde\lambda_i)
\end{equation}
where $\hat\sigma^{2}$ is given in \eqref{eq:NoisePower} and $\eta^{*}(\Tilde{\lambda}_i)$ is given in \eqref{eq:GeneralShrinker1}. The proof of optimality of this estimator is given in \cite[Sec.\,6]{donoho2018optimal}.
\end{enumerate}
 A pseudo-code to compute the estimator is stated in Algorithm~\ref{alg:Donohue}.
 \begin{algorithm}[h!]
\caption{Nonlinear Shrinkage Algorithm for Spiked Covariance Matrix Estimation}\label{alg:Donohue}
\begin{algorithmic}[1]
\State Evaluate the eigenvalue decomposition of sample covariance matrix $\SampleCov_{\Datasamples}$ as done in \eqref{eq:EigenDecompSample} for a given number of data samples $\Datasamples$.
\State Compute the noise power $\hat{\sigma}^2$ by \eqref{eq:NoisePower}.
\State Compute the eigenvalues $\EstEig_{i,\Datasamples}$ of the estimator as in \eqref{eq:EstEig}.
\State Using eigenvalues computed in Step 3, the estimated covariance matrix $\Est_{\Datasamples}$ is given by \eqref{eq:EigenDecompEst}.  
\end{algorithmic}
\end{algorithm}
\subsection{Convergence of Eigenvalues of the Proposed Estimator}
\label{Sec:Thm}

Although we are dealing with finite $p$ and $\Datasamples$, in Theorem\,\ref{Thm:NonLinBias} we state that the spiked eigenvalues converge almost surely to a constant and satisfy a Central Limit Theorem when both $p,\Datasamples\rightarrow\infty$, given that the number of spikes $r$ is fixed.

We assume the following for a covariance matrix $\NoiseClutter$ with dimension $p$:
\begin{compactenum}%[label=(A\arabic*)]
\item[A1.] Leading $r$ distinct eigenvalues  $\NoiseClutterEig_1,\,\NoiseClutterEig_2,\,\cdots,\,\NoiseClutterEig_r$ with multiplicity 1 and lower bounded by $1+\sqrt{\gamma}$.
\item[A2.] Eigenvalues $\NoiseClutterEig_{r+1}=1,\cdots,\NoiseClutterEig_{p}=1$.
\end{compactenum}
%\vspace{0.3cm}
\begin{theorem}
\label{Thm:NonLinBias}
Consider the estimator $\Est_{\Datasamples}$ of dimension $p$ with eigenvalues $(\EstEig_{1,\Datasamples},\,\EstEig_{2,\Datasamples},\,\cdots,\,\EstEig_{p,\Datasamples})$ that estimates the spiked covariance matrix $\NoiseClutter$ satisfying the assumptions (A1) and (A2). Assume $p/n\rightarrow\gamma\in(0,1),\,p,n\rightarrow\infty$. Then   $\EstEig_{i,\Datasamples}$'s,$\,\,1\leq i\leq r$ satisfy 
\begin{equation}
\label{eq:evLimit}
    \EstEig_{i,\Datasamples}\xrightarrow{a.s.}\nonLinFunc^{*}(\beta_i)
\end{equation}
Additionally, if $p/\Datasamples-\gamma=o(\Datasamples^{-1/2})$, then
\begin{equation}
\label{eq:NonlinearCLT}
    \sqrt{\Datasamples}(\EstEig_{i,n}-\nonLinFunc^{*}(\beta_i))\xrightarrow{d}\mathcal{N}(0,\alpha^2_i\,(\nonLinFunc^{'}(\beta_i))^2)
\end{equation}
where $\nonLinFunc^{*}(\cdot)$ is given by \eqref{eq:GeneralShrinker1} and $\nonLinFunc(\cdot)$ is given by \eqref{eq:Shrinker1}. $\beta_i=\NoiseClutterEig_i+\frac{\gamma\,\NoiseClutterEig_i}{\NoiseClutterEig_i-1}$,  $\alpha^2_i=~2\,\NoiseClutterEig_i^2\left(1-\frac{\gamma}{(\NoiseClutterEig_{i}-1)^2}\right)$ and $\gamma$ is given \eqref{eq:Kolmogorov}.
\end{theorem}
\textbf{Proof}: Almost sure convergence can be proved by applying Continuous Mapping Theorem on \cite[Thm.\,2]{paul2007asymptotics} with function $\nonLinFunc$. The in-distribution convergence can be proved by applying the delta method on \cite[Thm.\,3]{paul2007asymptotics} with function $\nonLinFunc$.\hfill$\blacksquare$

\begin{figure}
    \centering
    \includegraphics[width=\linewidth]{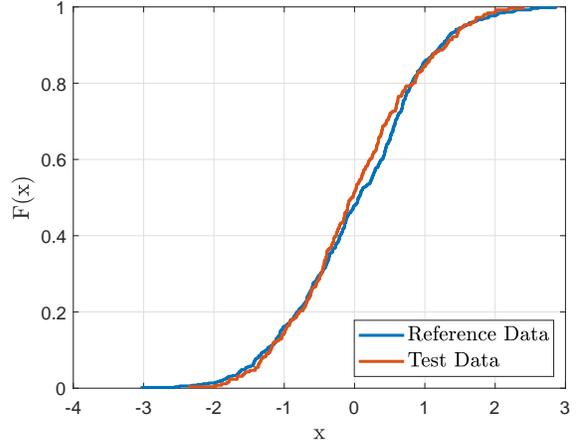}
    \caption{The double version of the K-S test with reference data from the prescribed distribution \eqref{eq:NonlinearCLT} and the test data from the Challenge Dataset verifies that the Challenge Dataset satisfies the assumptions for Thm.\ref{Thm:NonLinBias}  with $p$-value of 0.8390.}
    \label{fig:CDFPlot}
\end{figure}

We empirically showed the validity of assumptions of Theorem~\ref{Thm:NonLinBias} for the Challenge Dataset using the double version of the Kolmogorov-Smirnov~(K-S) test with significance level $5\%$ and 1024 Monte Carlo simulations. The
reference data was generated from the prescribed distribution in 
 \eqref{eq:NonlinearCLT}
and the test data was generated from the Challenge Dataset. The CDF plot in Fig.\ref{fig:CDFPlot}
for the test and reference data reveals that the Challenge Dataset satisfies the assumptions for Theorem~\ref{Thm:NonLinBias}.
%\vspace{-0.3cm}
\subsection*{Computation Cost}
Algorithm~\ref{alg:Donohue} does not require prior knowledge of the number of spikes. Step (2) and Step (3) in Algorithm~\ref{alg:Donohue} determine the eigenvalues that are above the noise floor. The computation cost of the algorithm is given below:
\begin{enumerate}
    \item The eigenvalue decomposition requires $\BigO(p^3)$ real-valued multiplications.
    \item The noise power estimation, step (2), is a median finding algorithm that requires $O(p)$ real-valued multiplications.
    \item The nonlinear shrinkage, step (3), requires $\BigO(r)$ real-valued multiplications, $r$ being the rank of the clutter covariance matrix.
\end{enumerate}
We compare algorithm \ref{alg:Donohue} to the RCML-EL algorithm  whose computational cost is given as follows:
\begin{enumerate}
    \item The eigenvalue decomposition step takes $\BigO(p^3)$ real valued multiplications. 
    \item The joint noise and rank estimation step takes $\BigO(p^2)$ real valued multiplications.
\end{enumerate}
The difference is in the noise and rank estimation step; Algorithm \ref{alg:Donohue} takes $\BigO(p)$ real-valued multiplications and the RCML-EL algorithm takes $\BigO(p^2)$. This will be demonstrated in Sec.~\ref{Sec:Simulations}-\ref{Sec:Challenge} empirically.

\subsection{Bounds for $\rho$}
\label{Sec:rhobound}
In this section, we derive the lower and upper bounds for the normalized SCNR($\rho$) using results from \cite{donoho2018optimal1}.

We rewrite $\rho$ from \eqref{eq:rho} 
\begin{eqnarray}\label{eq:raw_form}
\rho&=& \frac{\Vert \mathbf{x}\Vert_{2}^{2}}{(\mathbf{x}^H\Est^{-\frac{1}{2}}\NoiseClutter\Est^{-\frac{1}{2}}\mathbf{x})(\mathbf{x}^H\Est^{\frac{1}{2}}\NoiseClutter^{-1}{\Est}^{\frac{1}{2}}\mathbf{x})},
\end{eqnarray}
where  $\mathbf{x} = \hat{\NoiseClutter}^{-\frac{1}{2}}\chanSig$.
Without loss of generality, assume $\Vert \mathbf{x}\Vert=1$. We use the matrix version of Kantorovich's inequality to bound the denominator. 
For a positive semi-definite matrix, $\textbf{A}$, and a unit vector $\textbf{x}$, $\|\textbf{x}\|_2=1$,
$
(\textbf{x}^H\mathbf{Ax})(\textbf{x}^H\textbf{A}^{-1}\textbf{x})\leq\frac{1}{4}(\kappa(\textbf{A})+\frac{1}{\kappa(\textbf{A})}+2), 
$
where $\kappa(\textbf{A})$ is the condition number of the matrix $A$.
By Cauchy-Schwartz Inequality $(\textbf{x}^H\mathbf{Ax})(\textbf{x}^H\textbf{A}^{-1}\textbf{x})\geq 1$.
We lower bound $\rho$ by:
\begin{equation}
\label{eq:rhobound}
\rho\geq\frac{1}{\frac{1}{4}(\kappa(\textbf{A})+\frac{1}{\kappa(\textbf{A})}+2)}\end{equation}
where $\textbf{A}=\Est^{-\frac{1}{2}}\NoiseClutter\Est^{-\frac{1}{2}}$. From \cite{donoho2018optimal1}, we have 
\begin{equation} \label{eq:optpivot}
 \kappa(\textbf{A}) = \frac{ \max \left[ 1,\max_{1 \leq i \leq r}  \nu_+(\NoiseClutterEig_{i}^{\ast},\nonLinFunc_{i}) \right] }{\min \left[  1, \min_{1\leq i \leq r}  \nu_-(\NoiseClutterEig_{i}^{\ast},\nonLinFunc_{i}) \right]}
\end{equation}
where \[
   \nu_\pm(\NoiseClutterEig_{i}^{\ast},\nonLinFunc_{i}) = T/2  \pm \sqrt{T^2/4-D}
\]
\[
  D= {\nonLinFunc^{\text{St}}}/{\NoiseClutterEig^{\ast}},\quad
  T = (\frac{s^2 + {\nonLinFunc^{\text{St}}}c^2}{\NoiseClutterEig^\ast}+c^2+ {\nonLinFunc^{\text{St}}} s^2)\]
$\nonLinFunc^{\text{St}}$ defined in \eqref{eq:Shrinker1} and $\NoiseClutterEig_{i}^{\ast}=\NoiseClutterEig_i/\sigma^2$.

In Sec.~\ref{Sec:Simulations} we shall demonstrate that the proposed algorithm performs within the derived bounds.
\subsection{Performance similarity between Proposed Algorithm and RCML-EL Algorithm}
\label{Sec:Similar}
In this section, we show that the proposed algorithm and the RCML-EL algorithm will give similar SCNR performance.

The optimization problem for clutter plus noise covariance matrix estimation assuming noise power to be unity defined in \cite[(35)]{kang2014rank} is:
\begin{equation}
\label{eq:OptKang}
\begin{aligned}
\min_{\boldsymbol{\EstEig}} \quad & \textbf{d}^{T}\,\boldsymbol{\EstEig}-\mathbf{1}^{T}\,\log{\boldsymbol{\EstEig}}\\
\textrm{s.t.} \quad & \textbf{F}\,\boldsymbol{\EstEig}\preceq\textbf{g}\\
  &\textbf{E}\,\boldsymbol{\EstEig}=\textbf{h}\\
\end{aligned}
\end{equation}
where $d_{i}=\SampleCovEig_{i}$, recall from Sec.~\ref{Sec:SignalModel} that $\EstEig$ is eigenvalue of the estimator and $\SampleCovEig$ is the eigenvalue of the sample covariance matrix. \[\textbf{F}=\begin{bmatrix}\textbf{U}^{T}&\hspace{-0.3cm}-\textbf{I}_{p\times p}&\textbf{I}_{p\times p} 
\end{bmatrix}^{T}\in\real^{3p\times p}\]\[\textbf{g}=\begin{bmatrix}\mathbf{0}_{p\times 1}^{T}&\hspace{-0.3cm}-\boldsymbol{\epsilon}_{p\times 1}^{T}&\hspace{-0.2cm}\mathbf{1}_{p\times 1}^{T}
\end{bmatrix}^{T}\in\real^{3p\times 1}\] $\boldsymbol{\epsilon}_{p\times 1}=[\epsilon,\ldots,\epsilon]_{p\times 1},\, \epsilon>0$, 
\[\textbf{U}=\begin{bmatrix}
1&-1&0&0&\ldots&0\\
0&1&-1&0&\ldots&0\\
\vdots&\ddots&\ddots&\ddots&\vdots&\vdots\\
0&\ldots&\ldots&\ldots&1&-1
\end{bmatrix}\in \real^{p\times p}\] 
\[\textbf{E}=\begin{bmatrix}\mathbf{0}_{r\times r}&\mathbf{0}_{r\times p-r}\\
\mathbf{0}_{(p-r)\times r}&\textbf{I}_{p-r}\end{bmatrix}\in \real^{p\times p}\]\[\textbf{h}=\begin{bmatrix}0,\,0,\,\cdots,0_r,1,\,1,\cdots,1\end{bmatrix}^{T}\in \real^{p\times 1}.\] The first constraint in \eqref{eq:OptKang} enforces $\EstEig_i$ to be positive in descending order and the second constraint enforces the last $p-r$ eigenvalues of the estimator to be equal. These constraints enforce a spiked covariance matrix structure on the estimator, stated in Definition \ref{Def:SpikedCovariance}, in a high dimensional setting. In \cite{donoho2018optimal}, the optimization problem for estimating leading $r$ eigenvalues for Stein loss under a spiked covariance model is given by:
\begin{equation}
\label{eq:OptDon}
    \begin{aligned}
\min_{\EstEig_i,\,1\leq i\leq r} \quad & a_i\,\EstEig_i-b_i\,\log{\EstEig_i}+m_i
\end{aligned}
\end{equation}
where $a_i=c^2/g(\SampleCovEig_{i})+s^2$, $b_i=1$ and $m_i=1/g(\SampleCovEig_i)-1-a_i+\log(f(\SampleCovEig_i))$; $f(\cdot)$ is given in \eqref{eq:fx}, $g(\cdot)$ is given in \eqref{eq:gx}, $c(\cdot)$ and $s(\cdot)$ are given in \eqref{eq:Cosine}. 

Since the cost function of \eqref{eq:OptKang} is identical to \eqref{eq:OptDon} within a constant and the constraints of \eqref{eq:OptKang} are implicit to the optimization problem of \eqref{eq:OptDon}, the normalized SCNR, $\rho$ and the rank of the clutter covariance matrix, $r$, will be identical.

In Sec.~\ref{Sec:Simulations} we shall demonstrate that the SCNR performance for the proposed algorithm is identical to the RCML-EL algorithm with a reduced computation time.
\subsection{Low-Rank Adaptive Normalized Matched Filter Detection}
\label{Sec:Detection}
In this section, we shall use the Low-Rank Adaptive Normalized Matched Filter (LR-ANMF) detector for a rotation-invariant estimator as stated in \cite{vallet2019improved}. This detection scheme is independent of the eigenvalue shrinkage in \eqref{eq:GeneralShrinker1} and only depends on the eigenvectors of the sample covariance matrix in high dimensions. This detector is the same for both the proposed algorithm and the RCML-EL algorithm.

We have the following binary hypothesis for a single target:
\begin{align}
\label{eq:hypo}
    \begin{split}
       \nullhypo:\,& \vecrece\sim\mathcal{N}(0,\NoiseClutter) \\
       \althypo:\,&\vecrece\sim\mathcal{N}(h_t\vecsignal,\NoiseClutter)
    \end{split}
\end{align}
where $\nullhypo$ is the null hypothesis when no target is present and $\althypo$ is the alternate hypothesis when the target is present. The target signal $\vecsignal$ is defined in the same way as $\chanSig$ in \eqref{eq:rho} with a complex-valued amplitude $h_t$ and $\NoiseClutter$ is the clutter plus noise covariance matrix. The test statistics for the LR-ANMF with $n$ data samples, as stated in \cite{rangaswamy2004robust}, is 
\begin{equation}
\label{eq:TS}
    T_n=\frac{|\vecsignal\,\hat{\Pi}_n\,\vecrece|^2}{\|\hat{\Pi}_n\,\vecsignal\|^2}>\delta
\end{equation}
where \[\hat\Pi_n=\I-\sum_{i=1}^{r}\eigenvectorv_{i,n}\,\eigenvectorv_{i,n}^{H}\]  is a projection matrix constructed using the eigenvectors $\eigenvectorv_{i,n}$ defined in \eqref{eq:EigenDecompSample} corresponding to the $r$ spikes of the spiked covariance matrix and $\delta$ is the detection threshold. Recall from Sec.~\ref{Sec:SignalModel}-\ref{Sec:SpikedCovar} that $r$ is the rank of the clutter covariance matrix. The knowledge of noise power $\sigma^2$ is not impacting the detection since we are using assumption (A1) in Sec.~\ref{Sec:Donohue}-\ref{Sec:Thm} where $\sigma^2$ has already been estimated. We present the convergence theorems from \cite{vallet2019improved} that state the in-distribution convergence of the test statistics under $\nullhypo$ and $\althypo$.
\begin{theorem}[\cite{vallet2019improved}, Thm\,2] Under $\nullhypo$ and assumption (A1) the test statistics $T_n$ satisfies:
\begin{equation}
    T_n\xrightarrow{D}\chi^2(2)
\end{equation}
 where $\chi^2(2)$ is a chi-squared distribution with one complex degree of freedom. The probability of false alarm with detection threshold $\delta$ is:
 \begin{equation}
 \begin{split}
     P_{{FA}}&=\lim_{n\rightarrow{}\infty}\mathbb{P}(T_n>\delta|\nullhypo)=\int_{\delta}^{\infty}\exp({-x})\,dx\\ &=\exp{(-\delta)}
 \end{split}
 \end{equation}
\end{theorem}
\begin{theorem}[\cite{vallet2019improved}, Thm\,3] Under $\althypo$ and assumption (A1) the test statistics $T_n$ satisfies:
\begin{equation}
    \lim_{n\xrightarrow{}\infty}\sup_{x\in\real}\left|\mathbb{P}(T_n<x)-F\left(x;2,\Delta\right)\right|\xrightarrow{}0
\end{equation}
 where $F(x;2,\Delta)$ denotes the cumulative distribution of a non-central $\chi^2$ distribution with one complex degree of freedom and non-centrality parameter $\Delta$.
 \begin{equation}
     \Delta=\frac{2|h_t|^2}{\nu}
 \end{equation}
 $h_t$ is defined in \eqref{eq:hypo},
 \begin{align*}
     \begin{split}
         \nu=&\frac{1}{\|\Pi\vecsignal\|^2+\sum_{i=1}^{r}(1-c^2(\NoiseClutterEig_i))|\vecsignal^{H}\,\eigenvectoru_i|^2}\\&+\frac{\sum_{i=1}^{r}(\NoiseClutterEig_i-1)(1-c^2(\NoiseClutterEig_i))|\vecsignal^{H}\,\eigenvectoru_i|^2}{(\|\Pi\vecsignal\|^2+\sum_{i=1}^{r}(1-c^2(\NoiseClutterEig_i))|\vecsignal^{H}\,\eigenvectoru_i|^2)^2}
     \end{split}
 \end{align*}
 where $\Pi=\I-\sum_{i}^{r}\eigenvectoru_i\eigenvectoru_i^H$, $c(\cdot)$ is defined in \eqref{eq:Cosine}, $\eigenvectoru_{i}$ defined in \eqref{eq:EigenDecomTrue}, $r$ is rank of the clutter covariance matrix and $\vecsignal$ defined in \eqref{eq:hypo}. The corresponding target detection probability with detection threshold $\delta$ is:
 \begin{align}
 \begin{split}
     P_{{D}}&=\exp(-\Delta)\sum_{k=0}^{\infty}\frac{\Delta^k}{k!}\left[1-\frac{\int_{0}^{\delta}x^{k}\exp(-x)\,dx}{\Gamma(k+1)}\right]\\
     &=\lim_{n\rightarrow{}\infty}\mathbb{P}(T_n>\delta|\althypo)
     \end{split}
 \end{align}
 where $\Gamma(\cdot)$ is the gamma function.
\end{theorem}
 In Sec.~\ref{Sec:Simulations}-\ref{Sec:DetectionNumerical},  we compute the detection probabilities for different false alarm probabilities over a range of signal-to-noise ratios (SNR), we empirically evaluate the robustness of the detector for detecting a single target in the presence of multiple targets that act as contaminating clutter discretes. Contaminating clutter discretes are additional spikes that are present due to undesired targets. They are not part of clutter spikes and change the clutter covariance matrix rank from $r$ to $\hat{r}$.

To conclude this section, we proposed the nonlinear shrinkage-based rotation invariant estimator by using the sample covariance matrix. We stated the convergence of the spiked eigenvalues of the estimator. We stated the bounds for the normalized SCNR. The equivalence of the RCML-EL algorithm in a high dimensional setting to the proposed algorithm was established. A detector for target detection was stated for the proposed algorithm. 

\section{NUMERICAL EXAMPLES}
\label{Sec:Simulations}
We use a dataset generated using $\textnormal{RFView}^{\circledR}$ software that provides an accurate characterization of complex RF environments. It uses stochastic transfer functions \cite{gogineni2022high} to simulate the high-fidelity RF clutter encountered in practice. 

The dataset consists of a data cube in the time domain and is a multi-dimensional $N\times K\times \Datasamples$ matrix, where $N$ is the total number of (fast-time), $K$ is the slow time and $\Datasamples$ is the number of range gates in a specified coherent processing interval. For our case, we use the range gates as the number of data samples $\Datasamples$.

 In Sec.~\ref{Sec:Simulations}-\ref{Sec:Challenge}, we use the Challenge Dataset generated by RFView. We compare the performance of the Algorithm~\ref{alg:Donohue} against the RCML-EL based estimation algorithm given in \cite{kang2016expected}.
 We plot the normalized SCNR ($\rho$), stated in \eqref{eq:rho}, as a function of training data size $\Datasamples$, the normalized Doppler, and the normalized angle. For all the plots we are simulating in the regime where $\Datasamples=\BigO(p)$, i.e.,  $1/10<p/n<1$.  We also demonstrate the computation times for our proposed algorithm and the RCML-EL algorithm for various $\Datasamples$.
 
In Sec.~\ref{Sec:Simulations}-\ref{Sec:DetectionNumerical}, we compute the target detection probabilities over a range of false alarm probabilities and SCNR. In Sec.~\ref{Sec:Simulations}-\ref{Sec:ErrorVariance}, we compute the error variance of the minimum variance distortionless response beamformer using the proposed algorithm and compare it with the error variance corresponding to the RCML-EL algorithm and the true covariance matrix.

We only compare with the RCML-EL algorithm because it outperforms the Sample covariance matrix SMI, FML\,\cite{steiner2000fast}, Chen's algorithm\,\cite{chen2001development} and AIC\,\cite{akaike1974new}, as documented in \cite{kang2016expected} in all metrics. Since the theory underlying Theorem \ref{Thm:NonLinBias} holds only in the regime of $n>p$, no definitive statements can be made for the case of $n<p$. Therefore, the validity of the proposed algorithm is restricted to the case of $n>p$.

We simulated our results $\textnormal{Matlab}^{\circledR}$-R2021b on Windows-11 OS running on AMD Ryzen 7 5800H microprocessor with 16GB RAM.
\subsection{SCNR Performance}
\label{Sec:Challenge}
The Challenge dataset contains radar target and clutter returns generated by $\textnormal{RFView}^{\circledR}$. The scenario in Challenge Dataset has 4 targets and ground clutter containing buildings. This scenario involves an airborne monostatic radar flying over the Pacific Ocean near the coast of San Diego looking down for ground moving targets. The data spans several coherent processing intervals as the platform is moving with constant velocity along the coastline. In Table~\ref{Table:RadarPlatformLocation}-\ref{Table:Clutter2}, Appendix we state all the parameters used for this scenario.

The data set consists of a $32\times64\times2335$ data cube matrix which has the clutter impulse response over 32 channels with 64 pulses and 2335 data samples. We concatenate $8$ channels to get a clutter impulse response matrix of size $512\times2335$. We convolve the rows of the clutter impulse response matrix with a waveform of pulse length $1000$ to get a clutter return matrix of dimension $512\times 3334$. We add additive white Gaussian noise with zero-mean and variance $\sigma^2=5\times10^{-14}$ to the resulting clutter return matrix. The dimension of the clutter plus noise covariance matrix is $512\times 512$. We vary $\Datasamples$ in the multiples of $p$ till $n<3334$ to get the sample covariance matrix. For each plot, we use 1024 Monte Carlo simulations. For normalized Doppler, we fix the angle interval at $\frac{\pi}{180}$ and marginalize over it. For the normalized angle, we fix the Doppler interval at $\frac{\pi}{50}$ and marginalize over it. For both cases, we fix $\Datasamples=1024$.  

\begin{figure}[h!]
        \centering
        \includegraphics[width=\linewidth]{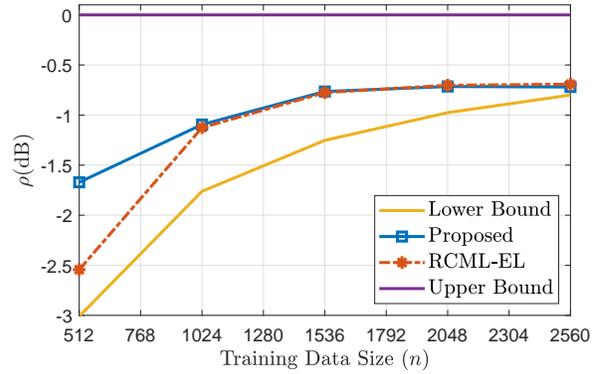}
        \caption{The Proposed algorithm has identical SCNR compared to the RCML-EL algorithm. SCNR for both estimators is within the derived bounds.}
        \label{fig:SDS12Compare}
    \end{figure}

\begin{figure}[h!]
        \centering
        \includegraphics[width=\linewidth]{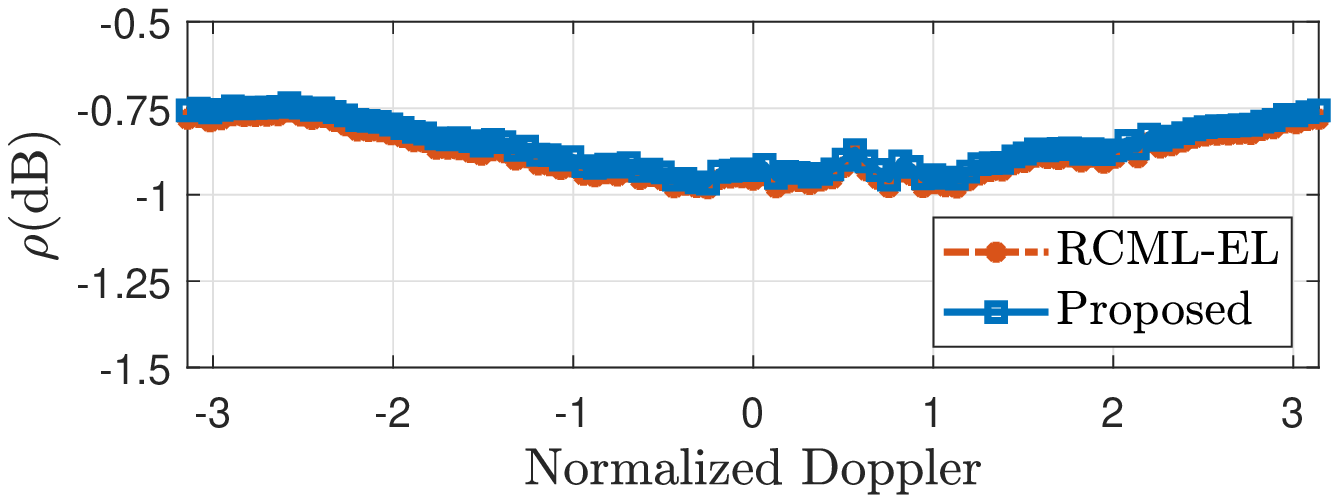}
        \caption{Algorithm \ref{alg:Donohue} has identical SCNR compared to the RCML-EL algorithm.}
        \label{fig:SDS12Doppler}
    \end{figure}

\begin{figure}[h!]
        \centering
        \includegraphics[width=\linewidth]{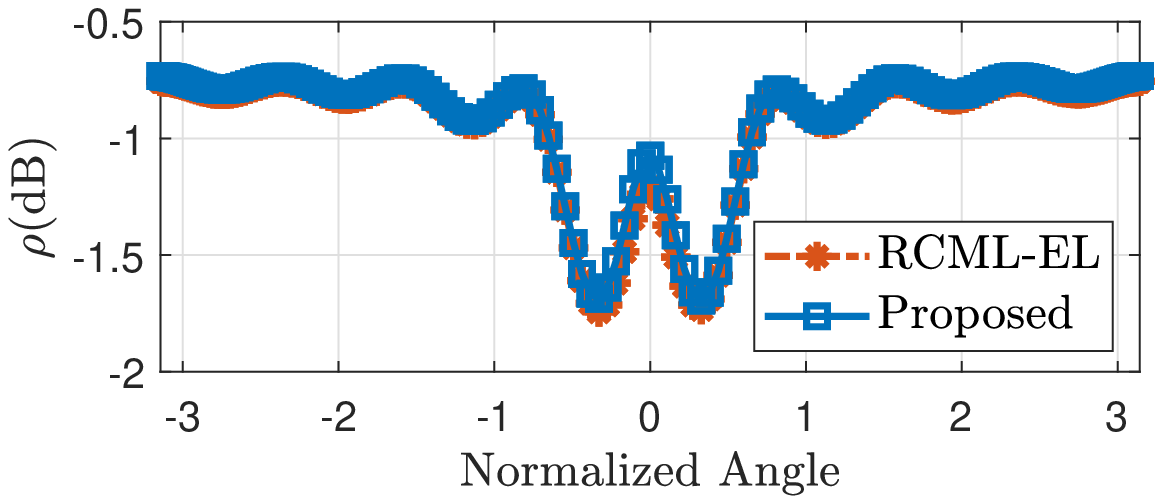}
        \caption{Algorithm \ref{alg:Donohue} has identical SCNR compared to the RCML-EL algorithm.}
        \label{fig:SDS12Angle}
    \end{figure}

Fig.\,\ref{fig:SDS12Compare} displays the average normalized SCNR vs. the number of data samples $\Datasamples$. The run-time for different $\Datasamples$ is given in Table \ref{table:SDS12}. Normalized SCNR vs. normalized Doppler is given in Fig.\,\ref{fig:SDS12Doppler} and normalized SCNR vs. normalized angle in Fig.\,\ref{fig:SDS12Angle}.
\begin{table}[h!]
\centering
\begin{tabular}{|p{2cm}|l|l|}
\hline
Training Data Size\,($\Datasamples$) & Proposed Algorithm & RCML-EL algorithm  \\\hline
512        & 0.012035s           & 0.107116s      \\\hline
1024       & 0.003046s           & 0.070876s      \\\hline
1536       & 0.002028s           & 0.053037s      \\\hline
2048       & 0.002026s           & 0.050819s      \\\hline
2560       & 0.002095s           & 0.050504s      \\\hline
\end{tabular}
\caption{Algorithm \ref{alg:Donohue} takes less computation time as compared to the RCML-EL algorithm for the Challenge Data set.}
\label{table:SDS12}
\end{table}

%\vspace{-0.3cm}
\subsection{Target Detection}
\label{Sec:DetectionNumerical}
In this section, we use results from Sec.~\ref{Sec:Donohue}-\ref{Sec:Detection} with the target having a Doppler of $f_d=0.2$ and an angle of $\theta=30^{\circ}$. We plot the target detection probability $P_D$ as we vary false alarm probability $P_{FA}$ from $10^{-5}$ to $10^{-1}$ in the multiples of 10 from SCNR=-10dB to 30dB. We use $n=1024$ data samples with 1024 Monte Carlo Simulations. 

The Challenge Dataset contains 4 targets. We consider detecting a single target with remaining targets constituting contaminating clutter discretes. These contaminating clutter discretes do not share the same characteristics as the target of interest so there is no self-target cancellation. Recall from Sec.~\ref{Sec:Donohue}-\ref{Sec:Detection} that the contaminating clutter discretes change the clutter rank to an unknown $\hat{r}$.  By introducing multiple targets as contaminating clutter discretes we demonstrate the robustness of the detector.

\begin{figure}[ht!]
    \centering
    \includegraphics[width=\linewidth]{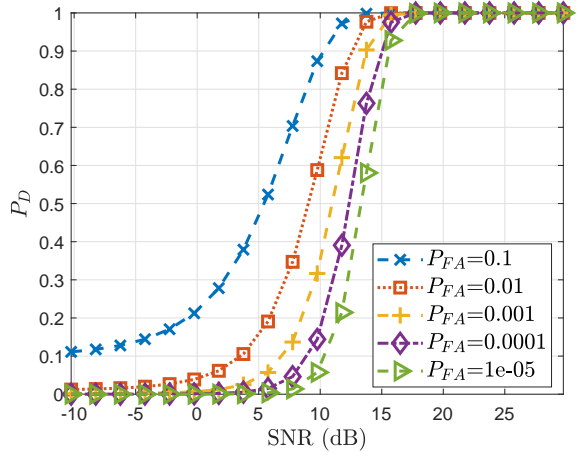}
    \caption{As $P_{FA}$ is decreased, higher SNR is required to get a fixed $P_D$ for a single target. The presence of other targets is not affecting the detection probabilities as they are projected into the null space of the target subspace. This empirically shows that the detector is robust to the presence of contaminating clutter discretes.}
    \label{fig:PdPfa}
\end{figure}

The detection probabilities are the same for both the RCML-EL algorithm and the proposed algorithm as the detector uses only the eigenvectors of the sample covariance matrix. The detection probabilities are illustrated in Fig.\ref{fig:PdPfa}. 
\subsection{Empirical Error Variance}
\label{Sec:ErrorVariance}
\begin{figure}
    \centering
    \includegraphics[width=\linewidth]{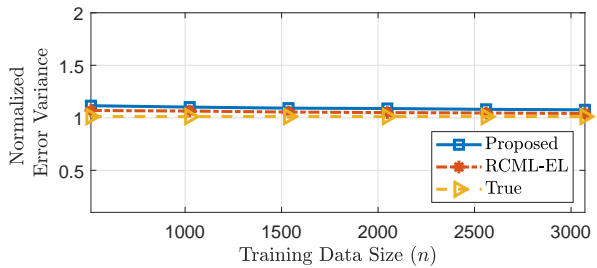}
    \caption{The normalized error variance due to the proposed algorithm and the RCML-EL algorithm is identical to that of the true covariance matrix. We normalize error variance in \eqref{eq:MSE} by the error variance of the true covariance matrix. }
    \label{fig:MSE}
\end{figure}

In this section, we empirically present the minimum variance distortionless beamformer (MVDR)  error variance due to the proposed algorithm with the error variance of the beamformer of the RCML-EL algorithm and the true covariance matrix. The error variance for the beamformer is 
\begin{equation}
\label{eq:MSE}
\text{Error Variance}=1/|\vecsignal^H\,\textbf{M}^{-1}\vecsignal|\end{equation} where $\textbf{M}=\Est_{\text{proposed}}$ for the proposed algorithm, $\Est_{\text{RCML-EL}}$ for the RCML-EL algorithm and $\NoiseClutter$ for the true covariance matrix, respectively. The target signal $\vecsignal$, as defined like $\chanSig$ in \eqref{eq:rho}, has Doppler $f_d=0.3$ and angle $\theta=30^{\circ}$. In Fig.\ref{fig:MSE}, the error variance for RCML-EL and the proposed algorithm is identical to the true covariance matrix. The error variance does not change as the training data size is increased since we are working in the asymptotic regime. This is due to the fact that in the asymptotic regime, the estimated covariance matrix converges to the true covariance matrix with probability 1. Hence, the error variance in \eqref{eq:MSE} merely becomes the reciprocal of the SNR from \eqref{eq:rho} when $\Est=\NoiseClutter$. 

To conclude this section, we demonstrated that with reduced covariance computation time, Algorithm \ref{alg:Donohue} gives identical SCNR performance compared to the EL-based covariance estimation algorithm within the proposed bounds. However, the noise computation step requires some pre-computed values of the medians of Marchenko Pastur distributions for various values of $\gamma$. This also makes our algorithm less robust to a sudden change in the parameters of the scenario as data samples can vary depending on the range swath. 

We demonstrated the target detection probabilities for different false alarm probabilities using the LR-ANMF detector. We empirically demonstrated the robustness with respect to contaminating clutter discretes in the Challenge Dataset. We empirically demonstrated that the error variance of the proposed algorithm is identical to the true covariance matrix.

\section*{CONCLUSION}
We exploited the spiked covariance structure for the clutter plus noise covariance matrix in a high dimensional setting and proposed a nonlinear shrinkage-based rotation invariant estimator. We stated the convergence of the spiked eigenvalues of the estimator. We demonstrated the reduced covariance computation times compared to the RCML-EL algorithm. Our proposed algorithm had identical SCNR performance compared to the RCML-EL algorithm. We employed the LR-ANMF detector for robust target detection and empirically showed that the error variance of the algorithm is identical to the true covariance matrix. 

Our proposed algorithm is a batch-wise algorithm. In future work it is worthwhile developing an adaptive version of the algorithm. We will also investigate other kinds of loss functions for various scenarios by introducing various constraints and deriving the concentration bounds for the proposed algorithm. The number of contaminating clutter discretes and their relative strength in the challenge dataset is not sufficient to provide a comprehensive analysis of the robustness feature of the LR-ANMF detector. This facet of the technique will be explored in more detail in the future. 

\appendix
\section{Challenge Dataset Parameters}
\label{Sec:Parameters}
In this section, we state the parameters we used for the Challenge Dataset in Table \ref{Table:RadarPlatformLocation}-\ref{Table:Clutter2}.
\begin{figure}[htbp!]
    \centering
    \includegraphics[scale=0.8]{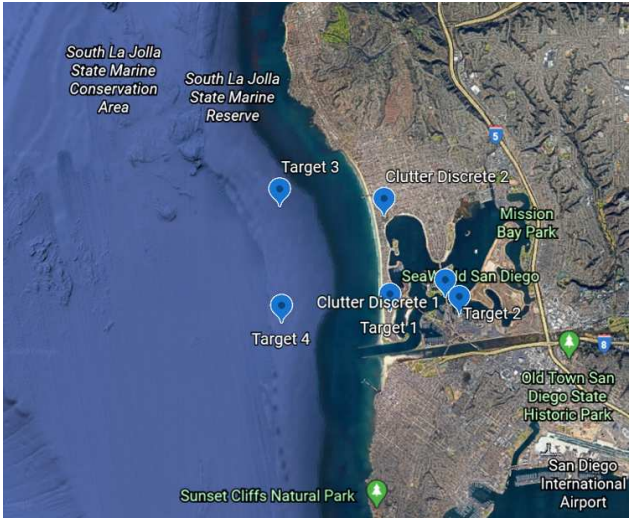}
    \caption{The Challenge Dataset scenario consists of $4$ targets and $2$ clutter discretes.}
    \label{scen1}
    \end{figure}
\begin{table}[h!]
\centering
\begin{tabular}{|p{4cm}|l|}
\hline
Latitude                                                    & 32.66 deg. N \\ \hline
Longitude                                                   & 118 deg. W   \\ \hline
Height                                                      & 6000 m       \\ \hline
Speed                                                       & 100 m/s      \\ \hline
Azimuth angle of   velocity vector (deg. w.r.t. true north) & 0 deg        \\ \hline
Elevation angle   of velocity vector (deg. w.r.t. horizon)  & 0 deg        \\ \hline
\end{tabular}
\caption{Radar Platform Location}
\label{Table:RadarPlatformLocation}
\end{table}
\begin{table}[h!]
\centering
\begin{tabular}{|p{4cm}|l|}
\hline
Number of Array   Elements (Horizontal Dimension)  & 32\\\hline
Number of Array   Elements (Vertical Dimension)    & 5\\\hline
Number of   Horizontal Spatial Channels (Receiver) & 32\\\hline
Number of   Vertical Spatial Channels (Receiver)   & 1\\\hline
Total Number of   Spatial Channels (Receiver)      & 32 \\\hline
Total Number of   Channels (Transmitter)           & 1 \\\hline
Transmit Antenna   Gain                            & 503.3509\\\hline
Receive Antenna   Gain                             & 15.7297 \\\hline
Center Frequency                                   & 10 GHz \\\hline
Array   Inter-Element Spacing                      & 0.015 m\\\hline
Number of   Coherent Processing Intervals (CPI)    & 30  \\\hline
Number of Pulses   per CPI                         & 64   \\\hline
Pulse Repetition   Frequency                       & 1 KHz   \\\hline
Radar Waveform                                     & Standard LFM\\\hline
Radar Waveform   Bandwidth                         & 10 MHz                \\\hline
Radar Waveform   Duty Factor                       & 0.1                    \\\hline
Sampling   Frequency                               & 10000000               \\\hline
Peak Transmit   Signal Power                       & 1000 Watts             \\\hline
Number of Range   Bins                             & 2334                  \\\hline
Size of Data   Cubes (for each CPI)                & 32 x 64 x 2334         \\\hline
Range Swath   Width                                & 20000 m                \\\hline
Radar Azimuth   Look Angle (Fixed)                 & 80.8321 deg            \\\hline
Radar Elevation   Look Angle (Fixed)               & -5.1364 deg            \\\hline
Clutter Scene   Size                               & 20Km x 20Km           \\\hline
Clutter Patch   Size                               & 20m x 20m              \\\hline
\end{tabular}
\caption{Monostatic Radar Parameters for the Challenge Dataset scenario.}
\label{Table:Monostatic Parameters}
\end{table}
\begin{table}[h!]
\centering
\label{Table:Target 1 parameters}
\begin{tabular}{|p{4cm}|l|}
\hline
Latitude                                                    & 32.7627 deg. N  \\ \hline
Longitude                                                   & 117.2524 deg. W \\ \hline
Height                                                      & 0 m             \\ \hline
Speed                                                       & 10 m/s          \\ \hline
Azimuth angle of   velocity vector (deg. w.r.t. true north) & 0               \\ \hline
Elevation angle   of velocity vector (deg. w.r.t. horizon)  & 0               \\ \hline
RCS                                                         & 40              \\ \hline
\end{tabular}
\caption{The first target is moving straight North on the ground on Ocean Front Walk near Mission Beach Park in San Diego.}
\end{table}
\begin{table}[h!]
\centering
\begin{tabular}{|p{4cm}|l|}
\hline
Latitude                                                    & 32.7668 deg. N  \\ \hline
Longitude                                                   & 117.2334 deg. W \\ \hline
Height                                                      & 0 m             \\ \hline
Speed                                                       & 20 m/s          \\ \hline
Azimuth angle of   velocity vector (deg. w.r.t. true north) & 0               \\ \hline
Elevation angle   of velocity vector (deg. w.r.t. horizon)  & 0               \\ \hline
RCS                                                         & 40              \\ \hline
\end{tabular}
\label{Table: Target 2 Parameters}
\caption{The second target is moving straight North on the ground on Ingraham Street near Sea World San Diego.}
\end{table}
\begin{table}[h!]
\centering
\begin{tabular}{|p{4cm}|l|}
\hline
Latitude                                                    & 32.793 deg. N  \\ \hline
Longitude                                                   & 117.283 deg. W \\ \hline
Height                                                      & 0 m            \\ \hline
Speed                                                       & 10 m/s         \\ \hline
Azimuth angle of   velocity vector (deg. w.r.t. true north) & 180            \\ \hline
Elevation angle   of velocity vector (deg. w.r.t. horizon)  & 0              \\ \hline
RCS                                                         & 20             \\ \hline
\end{tabular}
\label{Table: Target 3 Parameters}
\caption{The third target is moving south in the water off the coast of San Diego. It is a weaker target compared to the other targets in this simulation. }
\end{table}
\begin{table}[h!]
\centering
\begin{tabular}{|p{4cm}|l|}
\hline
Latitude                                                    & 32.763 deg. N  \\ \hline
Longitude                                                   & 117.283 deg. W \\ \hline
Height                                                      & 0 m            \\ \hline
Speed                                                       & 15 m/s         \\ \hline
Azimuth angle of   velocity vector (deg. w.r.t. true north) & 0              \\ \hline
Elevation angle   of velocity vector (deg. w.r.t. horizon)  & 0              \\ \hline
RCS                                                         & 30             \\ \hline
\end{tabular}
\label{Table: Target 4 Parameters}
\caption{The fourth target is moving north in the water off the coast of San Diego.}
\end{table}
\begin{table}[h!]
\centering
\begin{tabular}{|p{5cm}|l|}
\cline{1-2}
\multicolumn{1}{|l|}{Latitude}  & \multicolumn{1}{l|}{32.7665 deg. N}   \\ \cline{1-2}
\multicolumn{1}{|l|}{Longitude} & \multicolumn{1}{l|}{117.2305 deg. W}  \\ \cline{1-2}
\multicolumn{1}{|l|}{Height}    & \multicolumn{1}{l|}{6  m}              \\ \cline{1-2}
\multicolumn{1}{|l|}{Speed}     & \multicolumn{1}{l|}{0 m/s}             \\ \cline{1-2}
\multicolumn{1}{|l|}{RCS}       & \multicolumn{1}{l|}{50}              \\ \cline{1-2}
\end{tabular}
\caption{The first clutter object is an L shaped building inside Sea World San Diego}
\label{Table:Clutter1}
\end{table}
\begin{table}[h!]
\centering
\begin{tabular}{ll}
\cline{1-2}
\multicolumn{1}{|l|}{Latitude}  & \multicolumn{1}{l|}{32.7901 deg. N} \\ \cline{1-2}
\multicolumn{1}{|l|}{Longitude} & \multicolumn{1}{l|}{117.252 deg. W} \\ \cline{1-2}
\multicolumn{1}{|l|}{Height}    & \multicolumn{1}{l|}{6  m}           \\ \cline{1-2}
\multicolumn{1}{|l|}{Speed}     & \multicolumn{1}{l|}{0 m/s}         \\ \cline{1-2}
\multicolumn{1}{|l|}{RCS}       & \multicolumn{1}{l|}{50}           \\ \cline{1-2}
\end{tabular}
\caption{The second clutter object is a cube shaped building off Mission Blvd in San Diego}
\label{Table:Clutter2}
\end{table}

       \bibliographystyle{ieeetr}
        \bibliography{bibliography.bib}

\end{document}